\documentclass{svjour2}

\usepackage{graphicx}

\journalname{Int J Theor Phys}

\begin{document}

\title{Time variable $\Lambda$ and the accelerating Universe}

\titlerunning{Time variable $\Lambda$ ...}

\author{Utpal Mukhopadhyay \and Saibal Ray\footnote{Corresponding author} \and A. A. Usmani \and Partha Pratim Ghosh}
\authorrunning{Mukhopadhyay \and Ray \and Usmani \and  Ghosh}

\institute{Utpal Mukhopadhyay \at Satyabharati Vidyapith,
Nabapalli, North 24 Parganas, Kolkata 700 126, West Bengal, India
\email{utpal1739@gmail.com.} \and Saibal Ray \at Department of
Physics, Government College of Engineering \& Ceramic Technology,
Kolkata 700 010, West Bengal, India \email{saibal@iucaa.ernet.in}
\and A. A. Usmani \at Department of Physics, Aligarh Muslim
University, Aligarh 202 002, Uttar Pradesh, India
\email{anisul@iucaa.ernet.in} \and Partha Pratim Ghosh \at
Department of Physics, A. J. C. Bose Polytechnic, Berachampa,
North 24 Parganas, Devalaya 743 424, West Bengal, India
\email{parthapapai@gmail.com}}

\date{Received: date / Accepted: date}

\maketitle

\begin{abstract}
We perform a deductive study of accelerating Universe and focus on
the importance of variable time-dependent $\Lambda$ in the
Einstein's field equations under the phenomenological assumption,
$\Lambda =\alpha  H^2$ for the full physical range of $\alpha$.
The relevance of variable $\Lambda$ with regard to various key
issues like dark matter, dark energy, geometry of the field, age
of the Universe, deceleration parameter and barotropic equation of
state has been trivially addressed. The deceleration parameter and
the barotropic equation of state parameter obey a straight line
relationship for a flat Universe described by Friedmann and
Raychaudhuri equations. Both the parameters are found identical
for $\alpha = 1$.

\end{abstract}

\section{Introduction}
To account for the vast majority of mass in the observable
Universe and to explain its accelerated expansion, the physical
cosmology of today requires two outstanding concepts: (i) the
matter which does not interact with the electromagnetic force -
{\it dark matter}, and (ii) the hypothetical energy that tends to
increase the rate of expansion of the Universe - {\it dark
energy}.

Zwicky~\cite{Zwicky1937}, using Virial Theorem, had suggested for
a possible existence of dark matter long ago, which was later on
supported by the studies of rotation curves
\cite{Roberts1973,Einasto1974}, gravitational
lensing~\cite{Tyson1986,Subramanian1987}, CMB
anisotropy~\cite{Bode1995,Doroshkevich1996} and bullet
clusters~\cite{Clowe2007,Bradac2008}. The advent of inflationary
theory~\cite{Guth1981,Linde1982,Albrecht1982} led to the
convincing belief that $96\%$ of matter content of the Universe is
hidden mass constituted by $23\%$ dark matter and $73\%$ dark
energy~\cite{Amsler2008}. Dark matter plays a central role in
early Universe during structure formation and galaxy evolution
because of its nature to clump in sub-megaparsec scales. COBE and
CMB experiments suggest that baryonic dark matter is not more than
a small fraction of the total dark matter present in the Universe
\cite{Sahni2004}. More to it, observational constraint regarding
neutrino mass and relic neutrino
density~\cite{Elgaroy2002,Spergel2003} eliminate the possibility
of hot dark matter in favor of cold non-baryonic dark
matter~\cite{Sahni2004}. However, warm dark matter (WDM) which is
neither hot nor cold, rather an intermediate one between them, may
be considered as a dark matter candidate along with cold dark
matter. Thus, $\Lambda$-CWDM model consisting of a mixture of cold
and warm dark matter may be a possible alternative to
$\Lambda$-CDM model. A favourite choice of WDM is sterile
neutrinos~\cite{Dodelson1994,Shi1999,Dolgov2002,Abazajian2001a,Abazajian2001b,Asaka2005a,Asaka2005b}.
It has been shown~\cite{Dodelson1994} that oscillations of active
neutrinos in the primeval Universe can produce sterile neutrinos.
Boyarsky et al.~\cite{Boyarsky2008} have used recent results on
$\Lambda$-CWDM models to show that for each mass above 2 KeV, the
presence of at least one model of sterile neutrino can serve as
dark matter. It has also been observed~\cite{Gu2009} that sterile
neutrinos have the necessary features of a dark matter candidate.

Observational results for an accelerating
Universe~\cite{Perlmutter1998,Riess1998} favored the idea of an
accelerating agent that had been referred to as dark energy.  In
accordance with the data, the already introduced Standard Cold
Dark Matter (SCDM) model is found giving way to $\Lambda$-CDM
model that has an advantage of assuming a nearly scale-invariant
primordial perturbations and a Universe with no spatial curvature
as predicted by the Inflationary
theory~\cite{Guth1982,Hawking1982,Bardeen1982}. It is found in
good agreement with various observational
results~\cite{Tegmark2004}. In the $\Lambda$-CDM scenario the
present acceleration of the Universe cannot be a permanent feature
because, structure formation cannot proceed during acceleration if
it is assumed that dark energy is not coupled to dark matter.
Thus, the Universe must have undergone a decelerating phase prior
to the present accelerating phase~\cite{Amendola2003}.
Observational evidence~\cite{Riess2001} also supports this idea.
So, the deceleration parameter must have undergone a flip in sign
during cosmic evolution.

One of the candidates among the dark energy models is related to
dynamic $\Lambda$ (for an overview see~\cite{Overduin1998}
and~\cite{Sahni2000}). In fact, the concept of dark energy and the
physics of accelerating Universe appears to be inherent in the
$\Lambda$-term of Einstein's field equations. Herein, we perform a
study of accelerating Universe in context of the time-dependent
$\Lambda$ in the field equations and reveal the importance of
dynamic $\Lambda$ while addressing various key issues and some
known physical observable like dark matter, dark energy, geometry
of the field, deceleration parameter and its sign flip, age of the
Universe and equation of state parameter.

\section{The Einstein Field Equations and Their Solutions}
For the present purpose we consider the Einstein's field
equations,
\begin{equation}
  R^{ij} - \frac{1}{2}Rg^{ij} = -8\pi G\left[T^{ij} - \frac{\Lambda}{8\pi
  G}g^{ij}\right],
\end{equation}
which yield Friedmann equation
 \begin{equation}
\left(\frac{\dot a}{a}\right)^{2}+\frac{k}{a^2}=\frac{8\pi
G\rho}{3}+\frac{\Lambda}{3}, \label{fried}
\end{equation}
for the spherically symmetric FLRW metric. We also get
Raychaudhuri equation
\begin{equation}
\frac{\ddot a}{a}=-\frac{4\pi G}{3}(\rho+3p)+\frac{\Lambda}{3}
\label{akr}
 \end{equation}
in connection to the evolution scenario for expansion of null or
time-like geodesic congruences. Here, $\Lambda=\Lambda(t)$ is the
time-dependent function of the erstwhile {\it cosmological
constant} as introduced by Einstein, $a=a(t)$ is the scale factor
of the Universe and $k$ is the curvature constant.

The Raychaudhuri equation~(\ref{akr}) at once shows that for
$\rho+3p=0$, acceleration is initiated by the $\Lambda$-term only
that seems to relate $\Lambda$ with dark energy. It also shows
that for a positive $\Lambda$, the Universe may accelerate with
the condition $\rho+3p \leq 0$ i.e. $p$ is negative for a positive
$\rho$  with a definite contribution  of $\Lambda$ in the
acceleration. The placement of $\Lambda$ in the field equations
itself suggests for it to be a part of total energy momentum
$\tilde T^{ij} = T^{ij}-\Lambda g^{ij}/8\pi G$ of the Universe. In
fact, for a variable $\Lambda$, solutions of field equations are
possible only, if instead of $T^{ij}$, the total energy momentum
$\tilde T^{ij}$ is conserved~\cite{Vishwakarma2000} and in that
case the second term $(-\Lambda g^{ij}/8\pi G)$ of the above
equation acts as an additional source term in the field equations.
In the observational front, the data set coming from the Supernova
Legacy Survey (SNLS) during its first year of observation show
that dark energy behaves in the same manner as that of a
cosmological constant to a precision of $10\%$~\cite{Astier2006}.
Thus, $\Lambda$ being a time-dependent function, its  effect on
field equations in regard to the accelerating Universe is a
pertinent study.

Using Friedmann equation~(\ref{fried}) for  energy density
\begin{equation}
\rho=\frac{3}{8\pi
G}\left(\frac{k}{a^2}+H^2-\frac{\Lambda}{3}\right) \label{rho01}
\end{equation}
and the deceleration parameter
\begin{equation}
q=-\frac{a \ddot a}{\dot a^2}=-\frac{1}{H^2}\left(\frac{\ddot
a}{a}\right) \label{d1}
 \end{equation}
in Raychaudhuri equation~(\ref{akr}), one may arrive at  following
expression for the pressure
\begin{equation}
p=-\frac{1}{8\pi G}\left[\frac{k}{a^2}+(1-2q)H^2-\Lambda\right].
\label{p01}
\end{equation}

Now we Consider the phenomenological assumption $\Lambda \simeq
H^2$ ~\cite{Overduin1998,Sahni2000,Hsu2004,Ray2007}. Theoretical
and observational values of $\Lambda$ and $\rho$ for the present
day Universe, respectively $1-2 \times 10^{-35} {\rm
s}^{-2}$~\cite{Carmeli2002,Ray2007} and $4.5 - 18 \times 10^{-30}
{\rm gm cm}^{-3}$~\cite{Guth1997}, suggest a more general form
$\Lambda =\alpha H^2$ with $0.4 <\alpha <2.0$. As shown later, the
physical constraint  on $\alpha$ requires $0 < \alpha < 3$. In
view of this wide observational uncertainty in $\alpha$ with
regard to its relationship with the time variable $\Lambda$ should
thus be taken carefully while studying field equations for the
evolution of the Universe.

Under the same assumption $\Lambda=\alpha H^2$, for a flat
Universe, equation (\ref{rho01}) reads as
\begin{equation}
\rho_0=\frac{\Lambda_0}{8\pi G\alpha}(3-\alpha)=
\frac{\Lambda_0}{4\pi G} \mid_{\alpha=1} \label{rho02}
\end{equation}
giving thereby $\alpha=3\Lambda_0/(\Lambda_0 + 8\pi G\rho_0)$.
Similarly, equation (\ref{p01}) reads as
\begin{equation}
p_0=-\frac{\Lambda_0}{8\pi G\alpha}(1-\alpha-2q_0)=
\frac{q_0\Lambda_0}{4\pi G} \mid_{\alpha=1}.\label{p02}
\end{equation}
Here, subscript $0$ refers to the values of present day Universe.
Not ignoring the fact that the flat Universe would continue
evolving with variable $H$,  the present value, $H_0$, and other
related quantities ($\Lambda_0$, $p_0$, $\rho_0$ and $q_0$) may be
treated as dynamic and time variable.

Equations (\ref{rho02}) and (\ref{p02}) yield the barotropic
equation of state
\begin{equation}
\omega_0=\frac{p_0}{\rho_0}=-\frac{1-\alpha-2q_0}{3-\alpha}=q_0\mid_{\alpha=1}
\end{equation}
representing a straight line
\begin{equation}
q_0=\frac{3-\alpha}{2}\omega_0 + \frac{1-\alpha}{2} \label{d2}
\end{equation}
with $dq_0/d\omega_0=(3-\alpha)/2$. For $\alpha=1$, equation of
state parameter equals the deceleration parameter and therefore is
expected to obey the same physical conditions. More to it, as a
direct consequence of Friedmann and Raychaudhuri equations,  we
get while adding equations~(\ref{rho02}) and (\ref{p02})
\begin{equation}
p_0+\rho_0=\frac{\Lambda_0}{4\pi G\alpha}
(1+q_0)=\frac{\Lambda_0}{4\pi G} (1+q_0)\mid_{\alpha=1}.
\label{vacuu}
\end{equation}

It is well known that after the initial introduction of the
$\Lambda$ term by Einstein in his cosmological model, a number of
times that particular term has been accepted as well as rejected.
In the 1960's Zeldovich~\cite{Zeldovich1967,Zeldovich1968} revived
the same term as quantum fluctuations of the vacuum. He showed
that as a slow-rolling approximation of a scalar field in
Robertson-Walker space-time, the stress energy tensor takes the
form of a cosmological constant. This idea of Zeldovich played a
pivotal role on inflationary theory of Guth~\cite{Guth1981}. In
fact if we consider a scalar field $\Phi$ which obeys the
Klein-Gordon equation, then as slow-rolling approximation we have,
$p+\rho = \Phi^2 \approx 0$. The same situation has been obtained
in the present work via equation (11) for $q = -1$. This, for the
vacuum equation of state,
$p_0+\rho_0=0$~\cite{Davies1984,Hogan1984,Kaiser1984,Blome1984},
provides the condition of constant acceleration
($q_0=-1$)~\cite{Weinberg2008}. In general, the matter-energy
density being positive the counterpart negative pressure acts as a
repulsive agent and hence the vacuum equation of state has a deep
implication in the case of accelerating Universe scenario.
Obviously, for a collapsing Universe with positive $p$, we find
$q_0<-1$ and for an accelerating Universe we have $q_0>-1$. Its
higher limit may be positive depending upon the density. It is
evident from equation~(\ref{vacuu}) that the fate of the Universe
depends on $q_0$.

Bearing in mind the Hubble's law  and the assumption $\Lambda
=\alpha H^2$, we may directly arrive from equation~(\ref{d1}) at
$q_0= -1 -\dot\Lambda_0 /2\alpha H_0^3$.  This demonstrates that
for $\Lambda_0=0$ or for $\dot \Lambda_0/H_0^3=$ constant, the
Universe has been evolving through a constant acceleration as
indicated earlier~\cite{Weinberg2008}. It is interesting to note
here that Einstein initially obtained an expanding Universe with
$\Lambda=0$ (and hence to counteract the dynamical effects of
gravity, which would cause the matter-filled Universe to collapse,
he later on adopted a non-zero $\Lambda$ to obtain a static model)
while de Sitter obtained a similar expanding Universe with
constant $\Lambda$ and devoid of any ordinary matter (which
ultimately made Einstein to drop the cosmological constant from
his general relativistic field equations). We are also curious
about the situation $\Lambda_0=0$ which, by virtue of the
equation~(\ref{rho02}), makes the energy density to vanish. It
seems to  correspond to the special relativistic Universe of
Milne~\cite{Milne1935} under the zero-density limit of the
expanding FLRW metric with no cosmological
constant~\cite{Vishwakarma2000} but with $\Lambda_0=\alpha H_0^2$.

The available observational data for redshift and scale factor
have got flexibility, though limited, to distinguish between a
time varying and a constant equation of
state~\cite{Kujat2002,Bartelmann2005}. It, therefore, supports a
time variable $\omega_0$. Such a time variable $\omega_0$  has
been used in literature by many
authors~\cite{Nojiri2007,Brevik2007,Usmani2008} to predict various
physical observables of the Universe. Some useful limits on
$\omega_0$ was suggested by SNIa data, $-1.67 < \omega_0 <
-0.62$~\cite{Knop2003} whereas refined values come from combined
SNIa data with CMB anisotropy and galaxy clustering statistics
which are $-1.33 < \omega_0 < -0.79$~\cite{Tegmark2004}. Moreover,
inflation at an early stage scales the parameter $\omega_0$, which
combined with the above data  and dark energy constraint
($\omega_0>-1.0$) suggests a physical condition, $-0.46 > \omega_0
>-1.0$~\cite{Usmani2008}. In the light of our previous discussion,
$q_0$  too must obey the same physical conditions as $\omega_0$
but for $\alpha=1$. The experimental limits,
$-0.75<q_0<-0.48$~\cite{Cettoen08} does fall in this range, which
supports the view point that a variable $\omega_0$ provides
physical reason for a nonzero value of $\Lambda_0$ and for a
limited time variability for it. Moreover, the above ranges of the
values of $\omega_0$ and $q_0$ can be achieved if $\alpha$ is
constrained to lie in the range $0.5 < \alpha < 1.07$.

\begin{figure*}
\begin{center}
\vspace{0.5cm}\includegraphics[width=0.8\textwidth]{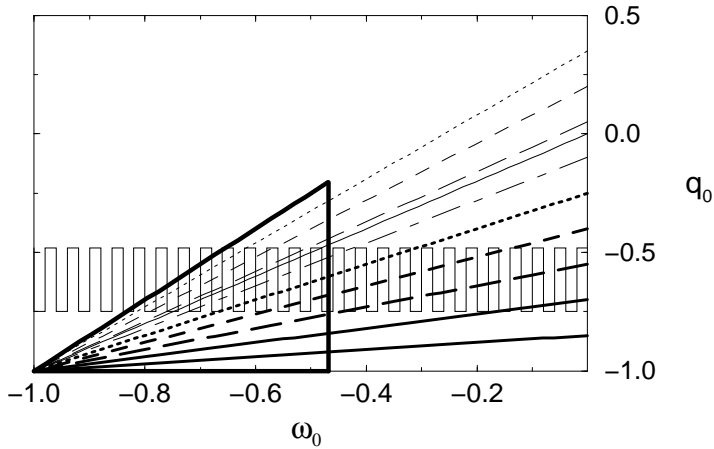}
\caption{Variation of $q_0$ with respect to $\omega_0$ as given by
equation~(\ref{d2}). The stair shaded area represents experimental
$q_0$ with uncertainty. The bold face triangle represents the
allowed region  due to physical conditions $0.0<\alpha<3.0$ and
$-0.46>\omega_0>-1.0$. The dotted, dashed, long-dashed, solid,
chain, thick dotted, thick dashed, thick long-dashed, thick solid
and thick chain curves, respectively represent the values of
$\alpha$ as $0.3$, $0.6$, $0.9$, $1.0$, $1.2$, $1.5$, $1.8$,
$2.1$, $2.4$ and $2.7$.\protect}\label{fig1}
\end{center}
\end{figure*}

For an accelerating Universe, $\Lambda_0$ must be positive, so is
$\alpha$. We plot the linear relationship~(\ref{d2}) for different
values of $\alpha$ within its physical range in Fig.~\ref{fig1}.
We choose $\omega_0$ from 0.0 to -1.0 covering the range $-0.46
>\omega_0 >-1.0$. To reproduce experimental $q_0$, we require
smaller values of $\alpha$ for smaller values of $\omega_0$. The
hypotenuse  and the base of the bold face triangle, respectively
represent $\alpha=0$ (or $\Lambda_0=0$) and $\alpha=3$, thereby
representing the full physical region. We may  compare $\alpha=0$
plot with other lines of the Figure representing positive values
of $\alpha$ (or respective values of $\Lambda_0$) to extract the
effect of $\Lambda_0$ on $q_0$. The thin solid line shows the
case, $\alpha=1$. We find $q_0$ negative within the bold face
triangular region demarcated by physical conditions
$0.0<\alpha<3.0$ and $-0.46>\omega_0>-1.0$. The values $\alpha >
9/4$ (or $\Omega_{\Lambda_0}>3/4$) lies below the experimental
$q_0$. Towards the smaller values of $\alpha$ ($\alpha<0.7$) and
higher values of $\omega_0$ ($-0.2<\omega_0<0.0$), $q_0$ is
observed to flip the sign in the accelerating epoch. Before the
present cosmic acceleration, which had started only recently (a
few Gyr earlier), the Universe was expanding with deceleration.
So, at the turnover stage (from deceleration to acceleration), the
deceleration parameter must have changed its sign. It is worth
noting here that $q_0$ and $\alpha$ (hence $\Lambda_0)$ determine
the fate of the Universe. One may arrive at the same conclusion
through recognizing the fact that density is directly proportional
to $\Lambda_0$. Hence, time variability of $\Lambda_0$ can not be
put aside while addressing the geometry of the Universe. One of
the predictions of the inflation theory is a flat Universe with a
large value of $\Lambda$ in the early stages of the Universe.
Thus, one may argue that it is the cosmological parameter that
determined the geometry of the Universe and made it flat during
inflation.

In this connection we mention that the equation (10) is nothing
but a special form of the general equation which can be obtained
easily from the equations (4) and (6) for flat Universe ($k=0$)
and using the ansatz $\Lambda=\alpha H^2$. It is also to be noted
here that the Fig.~\ref{fig1} demonstrates the evolution of the
Universe for a number of selected values of $\omega_0$. This
implies that general form of equation (10) will show the same
behavior in the $q$ versus $\omega$ plot. The equation (10) (or
it's general form) seems to be telling us that the cosmological
constant scales, via $H^2$, roughly as the `velocity' squared of
the Universe. So, equation (10) then will remain valid for a wide
range of cosmological epoch so far as the values of $\omega$ and
$q$ are concerned.

\begin{figure*}
\begin{center}
\vspace{0.5cm}\includegraphics[width=0.8\textwidth]{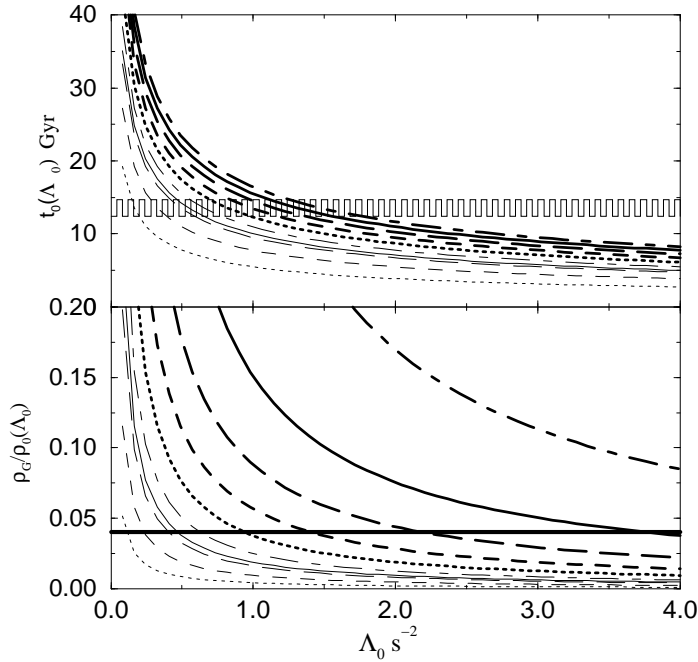}
\caption{The upper and lower panels show age of the Universe and
$\rho_G/\rho_0$ (representing hidden mass), respectively. The
stair shaded area shows observed age of Universe at present with
uncertainty. The horizontal line in the lower panel represents
observed hidden mass i.e. $96\%$. The description of various
curves for $\alpha$ is the same as in Fig.~\ref{fig1}.\protect }
 \label{fig2}
\end{center}
\end{figure*}

We now focus on the implications of $\Lambda_0$ on the hidden
mass. It should be mentioned here that a combination of Type Ia
supernova data and CMB anisotropy gives a best fit value for the
dark energy density parameter $\Omega_{\Lambda}$ as
$\Omega_{\Lambda} = {0.63}_{-0.23}^{+0.17}$~\cite{Efstathiou1998}.
This value of $\Omega_{\Lambda}$ yields, $\Lambda \approx 10^{-52}
m^{-2} \approx 10^{-35} s^{-2}$~\cite{Carmeli2002}. We may, from
equation~(\ref{rho02}), deduce that $\Lambda_0 \propto H_0^2$ is
equivalent to $\Lambda_0 \propto \rho_0$ as have been suggested
for other type of phenomenological
$\Lambda$-model~\cite{Vishwakarma2000}. It may also be shown that
the ratio $\rho_G/\rho_0$ gives a measure for the hidden mass of
the Universe. On the basis of the phenomenological model $\Lambda
\simeq H^2$~\cite{Ray2007}, the present density has been obtained
as $\rho_0=3.3 \times 10^{-30} {\rm gm cm}^{-3}$, which is close
to the lower limit of the value obtained by Guth~\cite{Guth1997}:
$4.5 - 18 \times 10^{-30} {\rm gm cm}^{-3}$. Therefore, knowing
the measured galactic mass density, $\rho_G=4.5 \times
10^{-31}{\rm gm cm}^{-3}$~\cite{Peebles1967}, one finds $\rho_G
\sim 0.025\rho_0 - 0.15\rho_0 $ suggesting thereby that the
galactic mass density is about $2.5\%$ - $15\%$ of the total mass
density of the present Universe. Hence, according to the present
model there is hidden mass ranging from $85\%$ - $97.5\%$, which
seemingly approves the value of Ref.~\cite{Amsler2008}. Variation
of $\rho_G/\rho_0$ is shown in the lower panel of Fig.~\ref{fig2}
with respect to $\Lambda_0$ and $\alpha$. Higher values of
$\Lambda_0$ and lower values of $\alpha$ demonstrate large missing
mass.

On the basis of the most recent observations
Weinberg~\cite{Weinberg2008} reports the age of the Universe as
$\approx 12.4-14.7$ Gyr. Dynamic $\Lambda$ model has also been
used to estimate the age~\cite{Overduin1998,Vishwakarma2002},
however it seems to suffer either from low-age like $5.4$
Gyr~\cite{Overduin1998} that is less than the estimated globular
cluster ages, $12.5 \pm 1.2$ Gyr~\cite{Cayrel2001} or it is as
high as $27.4 \pm 5.6$ Gyr~\cite{Vishwakarma2002}. With the
variations of $\alpha$ and $\Lambda_0$ we plot our calculations in
the upper panel of Fig.~\ref{fig2}. We notice that all the values
of $\alpha$ within its physical range reproduce observed age of
the Universe but with a suitable $\Lambda_0$. One requires,
smaller values of $\Lambda_0$, for the smaller values of $\alpha$
in order to be close with observation. This sheds light on the
important time variability of $H_0$ and its correlation to $H_0$,
$\Lambda_0 \propto H_0$.

\section{Conclusions}
The present work may be considered as a part of a series of papers
dealing with the behaviour of some phenomenological $\Lambda$-dark
energy models under specific assumptions on the physical
parameters $\Lambda$, $\omega$, G
etc.~\cite{Usmani2008,Ray2009,Mukhopadhyay2009a,Mukhopadhyay2009b,Mukhopadhyay2010}.
In those works various aspects of $\Lambda$ models have already
been explored. So, the present work has been carried out in the
same line and is guided by the motivation of revealing some new
features of dark energy in the framework of the present
accelerating Universe. In the present paper, it has been possible
to arrive at various interesting physical ideas of modern
cosmology through  simple considerations of time variability of
the observables of flat Universe, specially $\Lambda_0$. We have
considered the phenomenological assumption, $\Lambda_0 =\alpha
H_0^2$, for the full physical range of $\alpha$ and hence the
relevance of time variable $\Lambda$ with regard to various key
issues like dark matter, dark energy, geometry of the field, age
of the Universe, deceleration parameter and barotropic equation of
state  has been trivially addressed. Interestingly, the
deceleration parameter $q_0$ and the barotropic equation of state
parameter $\omega_0$ have been found to obey a straight line
relationship (equation~(\ref{d2})) with slope
$dq_0/d\omega_0=(3-\alpha)/2$ for a flat Universe described by
Friedmann and Raychaudhuri equations. For $\alpha=1$, both the
parameters, $\omega_0$ and $q_0$, are equal and hence are expected
to obey the same physical conditions. The assumption,
$\Lambda_0=\alpha H_0^2$, seems to represent the Milne Universe.
It has  been shown that within the physical limits of accelerating
Universe, $q_0$ may flip its sign towards lower end values of
$\alpha$ and higher end values of $\omega_0$.

\subsection*{Acknowledgments} The authors (SR \& AAU) would like
to express their gratitude to the authority of IUCAA, Pune for
providing them the Visiting Associateship under which a part of
this work was carried out. SR is personally grateful to Andrew
DeBenedictis, Simon Fraser University, Canada for helpful
discussion via electronic correspondence. We all are thankful to
the referee for valuable suggestions which have enabled us to
improve the manuscript substantially.

\end{document}